# Computerized Langmuir Probe Measurements in a Capacitively Coupled RF Discharge

M. El Shaer[1], A. S. Mohamed[1], A. Massoud[1], M. Mobasher[1], M. Wuttmann[2]

[1] Faculty of Engineering, Zagazig University, Zagazig, Egypt

[2] Institut Français d'Archéologie Orientale, Cairo, Egypt

A system of automated computerized Langmuir probe measurements is used in order to determine the plasma parameters in a plasma reactor constructed for cleaning of metallic artifacts by RF discharge. A compensated probe insures the suppression of the RF interference. The probe data are collected using a commercial data acquisition unit connected to a portable computer. The raw data are processed using wavelet transforms techniques which assures the de-noising of the probe signal without distortion of the probe I-V characteristic. The first and second derivatives of the I-V characteristic are determined. The measurement of the electron density spatial distribution in the inter-electrode distance indicates a flat density profile in the middle region of the discharge.

## 1. Introduction

RF hydrogen plasma discharge has been used in previous work for cleaning metallic artifacts from corrosion products and elimination of chlorides responsible for post corrosion, [1]. In this work we develop a Langmuir probe measuring system capable of the determination of the plasma temperature and density easily; those parameters are important during the plasma reactor operation as they influence the effectiveness of the treatment. The plasma reactor is constructed as a transportable arrangement able to operate in archaeological excavation areas; so special effort has been made in order to obtain automated probe measurements using low cost data acquisition unit connected to a portable computer. The data are automatically treated, so we can easily follow the variation of the plasma parameters as function of the main discharge parameters. The probe has a compensation circuit in order to minimize the RF interference with the measured probe signal.

## 2. Experimental setup
### 2.1. The plasma discharge

The discharge used in these measurements is a capacitive RF discharge working at 13.56 MHz as shown in Fig. 1. The discharge occurs between two circular movable electrodes of 10 cm diameter; the RF generator is connected to the powered electrode via a matching circuit, while the other electrode is earthed. Hydrogen gas is used at a pressure of 0.6 mbar measured by an absolute capacitive vacuum meter. The probe is movable and is inserted in the plasma between the two electrodes. Experiments are done at low RF power, up to around 50 W, but the measurements can be extended to 200 Watts which is the range of power normally applied in cleaning metallic artifacts.

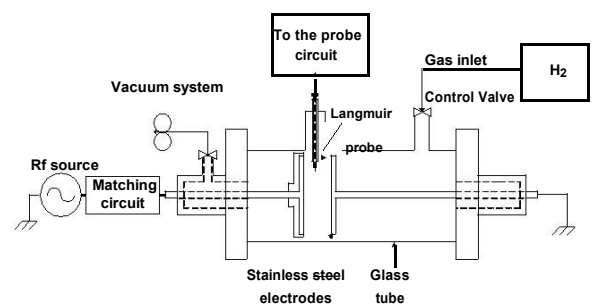

Fig. 1. Experimental setup

### 2.2. Probe construction

A compensated Langmuir probe is used to obtain the I-V characteristic, [2]. The probe tip is made of a tungsten wire of 0.3 mm diameter and 3 mm length.

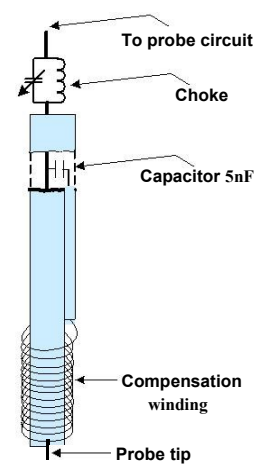

Fig. 2. RF Compensated Probe construction

The compensation circuit is achieved by winding 15 turns of tungsten wire of the same diameter around the probe; an RF tunable choke is connected just outside the probe and is calculated to be resonant at the fundamental RF discharge frequency.

This 15 turns winding which represents an external floating electrode is connected to the probe via a 5 nF ceramic capacitor, this connection is housed in the probe ceramic insulator as shown in Fig. 2. The idea of this noise suppression arrangement is to obtain by this winding a large area compared to the probe tip which intercepts the RF noise and feed it to the probe via the capacitor in order to be subtracted from the original probe signal. This auxiliary electrode with large impedance supplies only the RF voltage; the DC part is still supplied by the external power supply. The charge collected by this comparatively large probe is sufficient to drive the probe tip so that ($V_p$ - $V_s$) remains constant, where $V_p$ is the DC potential applied to the probe, and $V_s$ is the plasma space potential which can fluctuate at the RF frequency. A resonant RF choke with a variable capacitor is tuned at the fundamental RF frequency acting as an RF filter. Resonant circuits for higher harmonics are not used here.

### 2.3. Data acquisition system

We use a commercial USB data acquisition card (NI USB-6008). The card has a sampling rate of 10 kS/s, we use 2 input analogue channels for acquiring the voltage and current signals from the probe circuit and an output analogue channel for the control of the sawtooth generator delivering the applied voltage, as shown in Fig. 3. We overcome the slow rate in scanning by the superposition of more than 50 sweeps for each sawtooth voltage and current plots, and then averaging with suitable techniques.

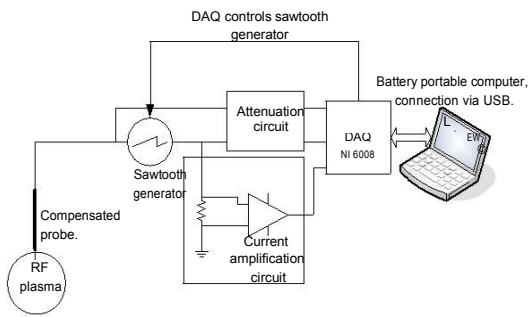

Fig. 3. Setup of probe circuit and data acquisition

The data are stored in a battery operated portable computer in order to suppress the noise through the AC line. For calculating the main plasma parameters, the data are processed using LabView programming which is a powerful instrument in data processing and analysis.

## 3. Experimental results
### 3.1. Processing of the probe data

An algorithm is applied for the denoising of the raw data of the Langmuir probe based on Daubechie's wavelet transforms (DWT) and bi-orthogonal wavelet transforms (BWT), which remove the noise from the raw data without losing important information, [3]. The advantage of applying the wavelet transforms is that it can minimize any manual intervention and improve the consistency and accuracy of the analysis by filtering the noise statistically.

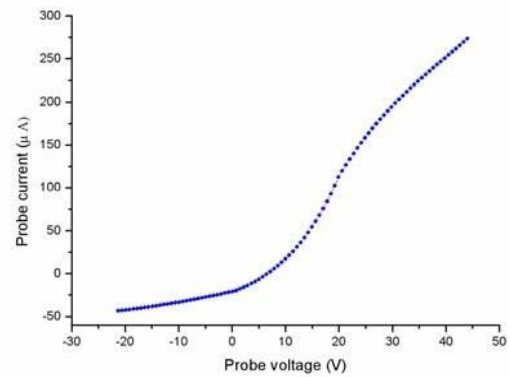

Fig. 4. Probe I-V characteristic after de-noising curve.

The I-V probe characteristic, shown in Fig. 4 is obtained after data processing using wavelet transforms. The DWT is used to filter out the noises from the signal in the electron and ion saturation regions because it has basis and scaling functions which are adequate for signals having a linear property. The high-pass filter of DWT can satisfactorily filter out the high frequency noise from any linear function, while the low pass filter preserves its linear state. For the transition region we use the BWT because it has functions suitable for noise reduction from signals with a curved property.

### 3.2. RF compensation

Fig. 4 is an example of compensated and filtered raw data. In order to insure that the I-V characteristic is properly compensated, we plot In Fig. 5 three I-V characteristics for the same discharge condition. First, without any compensation winding connected to the probe, second using a compensation winding, and third using besides the compensation winding, an RF choke tuned at the fundamental frequency.

The criterion for sufficient compensation is the increase of the floating potential to its maximum value of 7.3 V as seen in Fig. 5; this is achieved by tuning the variable capacitor in the RF choke. We can note that the compensation winding is more effective in the suppression of the RF interference than the RF choke which contribute in the compensation in a moderate approach.

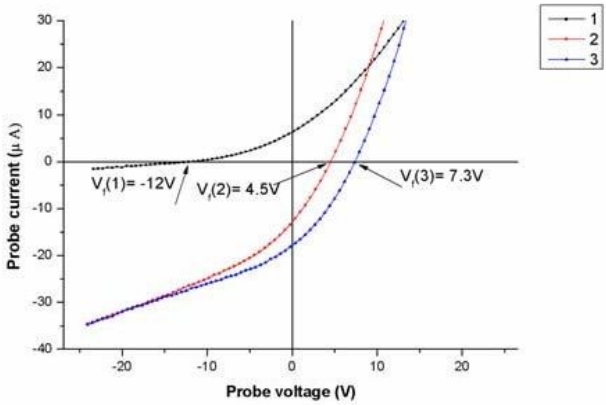

Fig. 5 Probe I-V characteristic, curve 1 is obtained without any compensation, curve 2 is with a compensation winding only and curve 3 is with compensation winding and RF tuning choke.

### 3.3 Determination of the plasma parameters

In Fig. 6, the first derivative of the probe electron current $I_e$ versus the probe voltage $V_p$ provides a peak value of 22 V, which corresponds to the space potential (plasma potential).

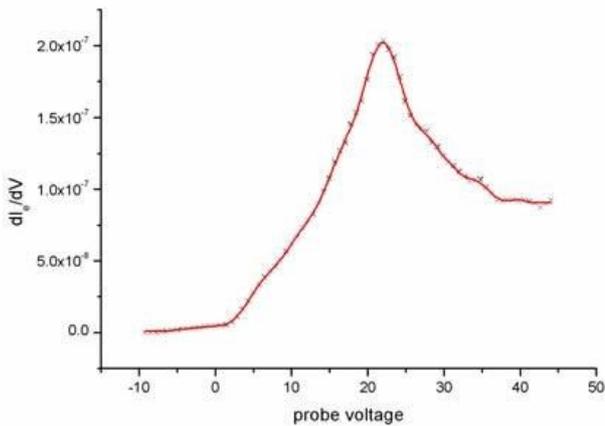

Fig. 6. First derivative the I-V characteristic

The second derivative of the electron current versus the probe voltage is shown in Fig. 7. For good compensation, the voltage difference between the maximum of the second derivative and its minimum should be as small as possible; this value is less than 10 volts, which is an indication of good compensation. From the second derivative we compute the electron energy distribution function (EEDF) shown in Fig. 8.

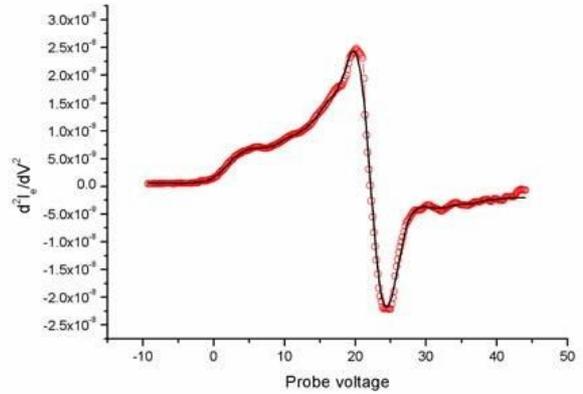

Fig. 7. Second derivative of the I-V characteristic, the calculated points from the measured data are superimposed with a smoothed curve

In Fig 8, the EEDF is plotted for different values of RF power. For low power the regime is collisional with a Druyvesteyn-like shape which is characterized by its convex shape, [4]. For higher power the EEDF seems to flatten, which is an indication of approaching a Maxwellian distribution with a collisionless heating regime.

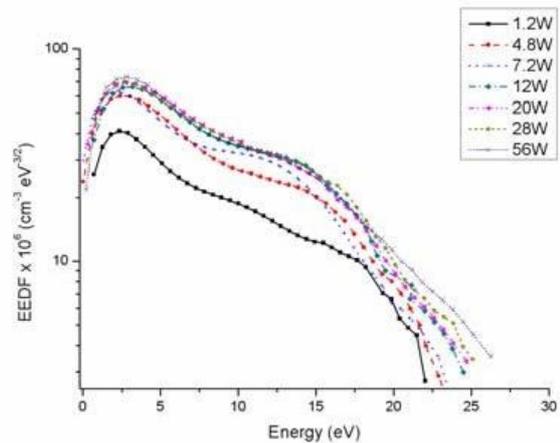

Fig. 8. Variation of the EEDF with RF power

The EEDF curve is used to compute the plasma density in the discharge by integration. The Spatial distribution of the electron density is shown in Fig. 9, which indicates a flattening of the electron density between the two electrodes; this curve is measured at a pressure of 0.6 mbar and a power of 20 W. For higher RF power the electron density increases with RF power and the electron temperature seems to have a value around 5 eV with a slight increase with RF power.

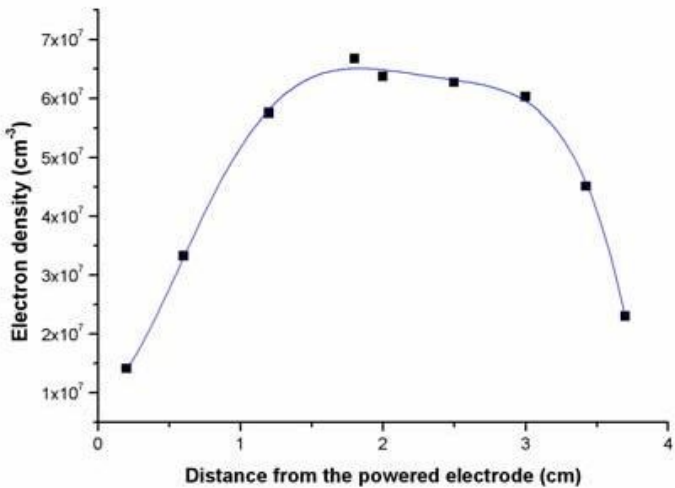

Fig. 9. Spatial distribution of density, the solid line is obtained using a polynomial fit

## 4. Conclusion

An automated Langmuir probe system is developed for measurements of the main plasma parameters in an RF discharge. Those parameters are important factors in plasma treatment during the cleaning of corroded artifacts in a plasma reactor operated with hydrogen. The use of a compensation coil has been proved to be effective in the suppression of the RF interference from the probe signal.